\documentclass[12pt, a4paper]{article}

    \usepackage{graphicx}

\begin{document}

    \title{The Role of the Exclusion Principle for Atoms to Stars: A Historical Account
    \footnote{Invited talk at the 12th Workshop on ``Nuclear Astrophysics'',
    March 22-27, 2004, Ringberg Castle, Germany.}}

    \author{Norbert Straumann\\
        Institute for Theoretical Physics University of Zurich,\\
        CH--8057 Zurich, Switzerland}

    \maketitle

    \begin{abstract}
    In a first historical part I shall give a detailed description of how Pauli discovered
    -- before the advent of the new quantum mechanics -- his exclusion principle. The
    second part is devoted to the insight and results that have been obtained in more
    recent times in our understanding of the stability of matter in bulk, both for ordinary
    matter (like stones) and self-gravitating bodies.
    \end{abstract}

    \section{Introduction}

    I have to apologize that my contribution will not be on a topic of current research. At
    this meeting in honor of Wolfgang Hillebrandt's 60th birthday it may not be out of
    place that my talk will have a historical accent. After all, Wolfgang has now become an
    elderly physicist with `his brilliant future (almost) behind him'.

    As a former student of Wolfgang Pauli, I was always interested in his way of doing
    science, which appears to me as an ideal of rare quality. Many of you rely in their
    daily work at least on two great discoveries of this man: the exclusion principle and
    the neutrino(s). In a contribution to Sommerfeld's 60th birthday in 1928, Pauli also
    developed the kinetic theory of particles that satisfy the exclusion principle \cite{P1}. All of
    you who work on core collapse induced supernova explosions use this theory, in one way or the
    other, in the study of the impact of the enormous neutrino pulse in supernova events.
    Wolfgang Hillebrandt and his coworkers have addressed this difficult task vigorously
    over many years. (I remember, for instance, very well the talk by Thomas Janka in
    Garching on his diploma thesis, that was devoted to a realistic treatment of the
    neutrino transport.)

    In the historical part of my talk, I shall sketch how Pauli arrived at the exclusion
    principle. At the time -- before the advent of the new quantum mechanics -- the exclusion principle was
    not at all on the horizon, because of two basic difficulties: (1) There were no general rules
    to translate a classical mechanical model into a coherent quantum theory, and (2) the spin
    degree of freedom was unknown. It is very impressive indeed how Pauli arrived at his principle
    on the basis of the fragile Bohr-Sommerfeld theory and the known spectroscopic
    material. The Pauli principle was not immediately accepted, although it explained many
    facts of atomic physics. In particular, Heisenberg's reaction was initially very
    critical, as I will document later. My historical discussion will end with Ehrenfest's
    opening laudation \cite{E} when Pauli received the Lorentz medal in 1931. This concluded with
    the words: \textit{``You must admit, Pauli, that if you would only partially repeal your
    prohibitions, you could relieve many of our practical worries, for example the traffic
    problem on our streets.''} According to Ehrenfest's assistant Casimir who was in the
    audience, Ehrenfest improvised something like this: \textit{``and you might also considerably
    reduce the expenditure for a beautiful, new, formal black suit''} (quoted in \cite{Enz},
    p.258).

    These remarks indicate the role of the exclusion principle for the stability of matter in
    bulk. A lot of insight and results on this central issue, both for ordinary matter
    (like stones) and self-gravitating bodies, have been obtained in more recent times, beginning with
    the work of Dyson and Lenard in 1967. Beside some qualitative remarks on a heuristic level, I
    intend to give in the second part of my talk a flavor of the deep insight, as well as of the
    concrete results, mathematical physicists have reached in this field over the last few decades.
    For further information, I highly recommend the review articles in Lieb's Selecta \cite{L1}.

    \newpage

    \section*{Part I. Wolfgang Pauli and the Exclusion Principle}

    Let me begin with a few biographical remarks. Pauli was born in 1900, the year of Planck's
    great discovery. During the high school years Wolfgang developed into an infant prodigy
    familiar with the mathematics and physics of his day.

    \section{Pauli's Student Time in Munich}

    Pauli's scientific career started when he went to Munich in autumn 1918 to study theoretical
    physics with Arnold Sommerfeld, who had created a ``nursery of theoretical physics''. Just
    before he left Vienna on 22 September he had submitted
    his first published paper, devoted to the energy components of the gravitational field in
    general relativity. As a 19-year-old student he then wrote two papers about the recent
    brilliant unification attempt of Hermann Weyl (which can be considered in many ways as the
    origin of modern gauge theories). In one of them he computed the perihelion
    motion of Mercury and the light deflection for a field action which was then preferred by
    Weyl.  From these first papers it becomes obvious that Pauli mastered the new field completely.

    Sommerfeld immediately recognized the extraordinary talent of
    Pauli and asked him to write a chapter on relativity in {\it Encyklop\"{a}die der
    mathemati\-schen Wissenschaften}. Pauli was
    in his third term when he began to write this article. Within less than one year he finished
    this demanding job, beside his other studies at the university. With this article \cite{P2} of 237 pages
    and almost 400 digested references Pauli established himself as a scientist of rare depth and
    surpassing synthetic and critical abilities. Einstein's reaction was very positive:
    \textit{``One wonders what to admire most, the psychological understanding
    for the development of ideas, the sureness of mathematical deduction, the profound physical
    insight, the capacity for lucid, systematic presentation, the knowledge of the literature,
    the complete treatment of the subject matter or the sureness of critical appraisal.''
    }Hermann Weyl was also astonished. Already on 10 May, 1919, he wrote to Pauli from Z\"{u}rich:
    \textit{`` I am extremely pleased to be able to welcome you as a collaborator. However, it is almost
    inconceivable to me how you could possibly have succeeded at so young an age to get hold of
    all the means of knowledge and to acquire the liberty of thought that is needed to assimilate
    the theory of relativity.''}

    Pauli studied at the University of Munich for six semesters. At the time when his Encyclopedia
    article appeared, he obtained his doctorate with a dissertation on the hydrogen
    molecule ion $H_2^+$ in the old Bohr-Sommerfeld theory. In it the limitations of the old
    quantum theory showed up.

    In the winter semester of 1921/22 Pauli was Max Born's assistant in G\"{o}ttingen. During
    this time the two collaborated on the systematic application of astronomical perturbation
    theory to atomic physics. Already on 29 November, 1921, Born wrote to Einstein: \textit{``Little
    Pauli is very stimulating: I will never have again such a good assistant.''} Well, Pauli's
    successor was Werner Heisenberg.

    \section{Discovery of the Exclusion Principle}

    Pauli's next stages were in Hamburg and Copenhagen. His work during these
    crucial years culminated with the proposal of his exclusion principle in December 1924.
    This was Pauli's most important contribution to physics, for which he received a belated
    Nobel Prize in 1945. Since this was made before the advent of the new quantum
    mechanics, I ask you to forget for a while what you know about quantum mechanics.

    The discovery story begins in fall 1922 in Copenhagen when Pauli began to concentrate his
    efforts on the problem of the anomalous Zeeman effect. He later recalled: \textit{`A colleague who met
    me strolling rather aimlessly in the beautiful streets of Copenhagen said to me in a friendly
    manner, ``You look very unhappy''; whereupon I answered fiercely, ``How can one look happy
    when he is thinking about the anomalous Zeeman effect?'' '.}

    In a Princeton address in 1946 \cite{P3}, Pauli tells us how he felt about the anomalous
    Zeeman effect in his early days:
    \begin{quote}

   \textit{ ``The anomalous type of splitting was on the one hand especially fruitful because it
    exhibited beautiful and simple laws, but on the other hand it was hardly
    understandable, since very general assumptions concerning the electron, using classical
    theory as well as quantum theory, always led to a simple triplet. A closer
    investigation of this problem left me with the feeling that it was even more
    unapproachable (...). I could not find a satisfactory solution at that time, but
    succeeded, however, in generalizing Land\'e's analysis for the simpler case (in many
    respects) of very strong magnetic fields. This early work was of decisive importance
    for the finding of the exclusion principle.''}

    \end{quote}

    \newpage

    \noindent {\bf Step 1: Zeeman effect for strong fields and Pauli's sum rule}

    \medskip

    \noindent
    I would like to show you now in some detail what Pauli did in his first step
    \cite{P4}. In doing this, I use `modern' (post-quantum mechanics) notations and
    first summarize the state of knowledge at the time when Pauli did his work.

    \begin{itemize}
    \item The energy levels of an atom determine the spectrum by
    {\it Bohr's rule}:
    \[
    E_2-E_1=h\,\nu. \]
    \item In spectroscopy some {\it
    quantum numbers} were already associated to energy levels, namely\footnote
    {In square brackets I give the historical notation.}:
    \begin{description}
    \item[$\triangleright$] $L~
    [=k-1]~,~~L=0,1,2,3,\ldots~~~(S,P,D,F,\ldots)$, \\
    our present day orbital angular momentum.
    \item[$\triangleright$] $S~ [=i-\frac 1 2]$: ~Each term belongs to a singlet or multiplet
    system, characterized by a {\it maximal} multiplicity ~ $2 S + 1~ (S=0, \frac 12, 1, \ldots
    )$, reached with increasing $L$. $S$ is our present day spin quantum number.
    \item[$\triangleright$] The various terms of a multiplet, having the \textit{same}
    $L$ and $S$, are distinguished by a quantum number $J$ [Sommerfeld's $j$], which takes
    the values:\\

    $\phantom{mmmmmm}J=L+S,~L+S-1,\ldots~~ L-S$~~ for~~$L \geq S$~,\\
    $\phantom{mmmmmm}J=S+L,~S+L-1,\ldots~~ S-L$~~ for~~$L < S$~.\\

    $J$ is our present day total angular momentum. The maximal multiplicity $2S+1$ is
    reached for $L\geq S$.
    \end{description}
    \item One knew the following {\it selection rules} (valid in most cases):
    \begin{eqnarray}
    &&L\;\longrightarrow\; L\pm 1~,\nonumber \\
    &&S\;\longrightarrow\;S~,\nonumber\\
    &&J\;\longrightarrow\;J+1, J, J-1~~~(0~\longrightarrow ~0{\rm
    ~forbidden}).\nonumber
    \end{eqnarray}
    \item For a given atomic number $Z$ ($Z-p$ if the atom is ionized  $p$ times) the
    following holds:
    \begin{eqnarray}
    &&Z~~{\rm even}~~~\longrightarrow~S, J\,:~~{\rm integer}~,\nonumber\\
    &&Z~~{\rm odd}~~~~\longrightarrow~S, J\,:~~{\rm  half~integer}~.~~~~~~~~~~~~~\nonumber
    \end{eqnarray}
    \item {\it Splitting in a magnetic field:}
    \begin{description}
    \item[$\triangleright$] Each term splits into $2J+1$ terms, distinguished by a
    quantum number $M$ taking the values $M=J, J-1, \ldots, -J$.
    \item[$\triangleright$] {\it Land\'e:} If the field is \textit{weak}, the terms are
    equidistant and their deviation from the unperturbed term is ~ $\Delta E_M = M \,\cdot\,g (\mu_0 B)$~,
    where ~$\mu_0 = e\hbar/2mc$~ is the Bohr magneton (introduced by Pauli in 1920) and $g$ is
    Land\'e's $g$ factor:
    \[
    g\;=\;\frac 3 2 \;+\;\frac{S(S+1) - L(L+1)}{2J (J+1)}~; \]
    \item[$\triangleright$] {\it Selection rules} for Zeeman transitions:
    \begin{eqnarray}
    &&M\;\longrightarrow\;M\pm 1~~(\sigma-{\rm component}),\nonumber\\
    &&M\;\longrightarrow\;M~~~~~~~ (\pi-{\rm component}).~~~~~~~\nonumber
    \end{eqnarray}
    If the $g$ factors for the initial and final states are the same, we have the following
    triplet:

    \setlength{\unitlength}{1mm}
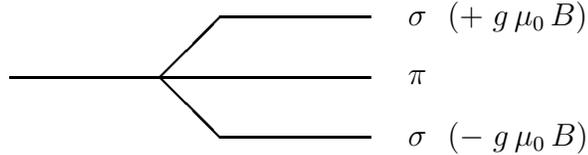
    \begin{picture}(150,40)(0,-20)
    \thicklines
    \put(20,0){\line(1,0){20}}
    \put(40,0){\line(1,1){8}}
    \put(40,0){\line(1,-1){8}}
    \put(48,8){\line(1,0){20}}
    \put(48,-8){\line(1,0){20}}
    \put(40,0){\line(68,0){28}}
    \put(73,8){\makebox(0,0)[s]{$\sigma~~~(+\;g\,\mu_0\,B)$}}
    \put(73,0){\makebox(0,0)[s]{$\pi$}}
    \put(73,-8){\makebox(0,0)[s]{$\sigma~~~(-\;g\,\mu_0\,B)$}}

    \end{picture}

    \end{description}
    \end{itemize}

    \noindent Pauli accepts these {\it empirical rules} as established, and proceeds to
    investigate the spectroscopic material for \textit{strong} fields. In a table he gives
    the energy splitting's $\Delta E$ as multiples of $\mu_0B$ and describes the result as follows:

    If two quantum numbers $M_L, M_S~ [= m_1,\mu]$ are introduced, whose sum is equal to  $M$,
    \[
    M\;=\;M_L+M_S~,~~~~~~~~~~~~~~~~\]
    and which take the values
    \[
    M_L\;=\;L, L-1, \ldots, -L~,~~~~~~~~~~~~~~~~~~~~~~~~~~~~~~
    \]
    \[
    M_S\;=\;\left\{ \matrix{ &\!\!\!\!\!\pm\;\frac 12 &\!\!\!\!\!\!\!\!{\rm
    for~ doublets~(alkali~ atoms)}\cr
         &0,\;\pm\; 1   &{\rm for~ triplets~(alkaline~ earths)}~,}\right.
    \]
    then the following simple formula holds for strong fields:
    \[
    \triangle E/\mu_0 B\;=\;M_L+2M_S\;=\;M+M_S~.
    \]
    Pauli {\it generalizes} this at once to arbitrary multiplets, assuming that the same
    formula holds, but that $M_S$ takes the values
    \[
    S, S-1, \ldots , - S.~~~~~~~~~~~~~~~~
    \]
    This generalization was at the time not experimentally tested.

    The selection rule for $M_S$ is: ~$M_S \to M_S$, hence the Zeeman effect is
    {\it normal for strong fields}. Thus the situation is simpler in this case.

     As the main point of the paper Pauli postulates a remarkable formal rule which
     allows him to derive Land\'e's whole set of  $g$ factors. Pauli's sum rule reads:

    \begin{quote}
    \textit{``The sum of the energies of all states of a multiplet belonging to  given values of
    $M$ and $L$ remains a linear function of $B$, when we pass from weak to strong fields.''}
    \end{quote}
    (In quantum mechanics this rule follows immediately
    \footnote{The sum in Pauli's rule is
    the trace of $\langle H_B\rangle$, $H_B=\mu_0B (J_3+S_3)$, where $\langle H_B\rangle$
    is the perturbation matrix for fixed $M$. This trace is obviously linear in $B$.}.)

    \medskip

    \noindent {\bf Special cases}:
    \begin{enumerate}
    \item[1)] $M=J=L+S$.
    Then there is only \textit{one } state, whose energy must be linear in $B$.
    \item[2)] If $M$ is chosen such that there are $2S+1$ states (the maximal possible
    number) then the arithmetic mean of their energies $\Delta E_M$ in strong magnetic
    fields is $M\mu_0B$. Hence Pauli's sum rule, which later became known as \textit{Pauli's
    g-permanence rule }, implies that the mean of all Land\'e factors is equal to
    1, for all $L\geq S$.
    \end{enumerate}

    \noindent Pauli now shows -- and he puts most weight to this -- that all factors $g$
    {\it can uniquely be calculated} from the energies for \textit{strong} fields. He shows
    this by applying the sum rule recursively for different values of $M$ with a given $L$.
    (This might serve as a nice exercise for students in a quantum mechanics course.)

    Pauli was very unhappy when he wrote this paper, which only later turned out to be
    important. In several letters he laments about his `unfortunate work on the anomalous Zeeman
    effect'. To Sommerfeld he wrote \footnote{\textit{ ``Ich habe mich sehr lange mit dem
    anomalen Zeemaneffekt geplagt, wobei ich oft auf
    Irrwege geriet und eine Unzahl von Annahmen pr\"ufte und dann wieder verwarf. Aber es
    wollte und wollte nicht stimmen! Dies ist mir bis jetzt einmal gr\"undlich
    danebengegangen! Eine Zeit lang war ich ganz verzweifelt ... ich habe das Ganze mit
    einer Tr\"ane im Augenwinkel geschrieben und habe davon wenig Freude.''}
     }:
    \begin{quote}
    \textit{``I have long vexed myself with the anomalous Zeeman effect and often lost my way. I
    considered and discarded untold assumptions. But it just wouldn't ever work out! In
    this I have miserably failed for once up to now! For a time I was quite desperate ... I
    have written all of this with a tear in the corner of my eyes and am anything but
    delighted.''}
    \end{quote}

    In the  final section of his paper he expresses very clearly why he believes that the
    presently known principles of quantum theory will not lead to an understanding of the
    anomalous Zeeman effect. Since I find it very difficult to preserve the characteristic
    style of Pauli's writing I quote only the German original:
    \begin{quote}
    \textit{``Eine befriedigende modellm\"assige Deutung der dargelegten Gesetz\-m\"assigkeiten,
    insbesondere der in diesem Pararaphen besprochenen formalen Regel ist uns nicht gelungen.
    Wie schon in der Einleitung erw\"ahnt, d\"urfte eine solche Deutung auf Grund der bisher
    bekannten Prinzipien der Quantentheorie kaum m\"oglich sein. Einerseits zeigt das Versagen des
    Larmorschen Theorems, dass die Beziehung zwischen dem mechanischen und dem magnetischen
    Moment eines Atoms nicht von so einfacher Art ist wie es die klassische Theorie
    fordert, indem das Biot-Savartsche Gesetz verlassen oder der mechanische Begriff des
    Impulsmomentes modifiziert werden muss. Anderseits bedeutet das Auftreten von
    halbzahligen Werten von $m$ und $j$ bereits eine grunds\"atzliche Durchbrechung des
    Rahmens der Quantentheorie der mehrfach periodisch\-en Systeme.''}
    \end{quote}

    After his return to Hamburg Pauli began to think about the closing of electronic
    shells. He was convinced that there must be a closer relation of this problem to the
    theory of multiplet structure. In his Nobel Prize lecture he writes:
    \begin{quote}
    \textit{ ``I therefore tried to examine again critically the simplest case, the doublet
    structure of the alkali spectra. According to the point of view then orthodox, which
    was also taken over by Bohr in his lectures in G\"ottingen, a non-vanishing angular
    momentum of the atomic core was supposed to be the cause of this doublet structure.''}
    \end{quote}

    In his next paper \cite{P5} Pauli rejected this `orthodox' point of view, and introduced instead
    a classically non-describable two-valuedness of the electron, now called the spin.

    \vspace{.5cm}
    \newpage
    \noindent {\bf Step 2: Two-valuedness of the electron}

    \medskip

    \noindent
    Let me show you in some detail how he arrived at this fundamental conclusion.
    First, he calculates the relativistic corrections upon the magnetic moment and the
    orbital angular momentum of electrons in the K-shell. For the ratio of the two he finds
    with simple classical arguments
    \[ \frac{|\vec{M}|}{|\vec{L}|}\;=\;\frac{e}{2mc}~\big\langle \left(
    1-v^2/c^2\right)^{1/2}\big\rangle_{\rm time}~.\]

    According to the virial theorem the time average on the right must be equal to the total
    energy of the electron in units of $mc^2$. For the latter Pauli uses Sommerfeld's
    relativistic formula and finds for the K-shell $(L=0, n=1)$ the value
    $(1-\alpha^2Z^2)^{1/2}$ for the relativistic correction factor $( \simeq 1
    -\,\frac12~\alpha^2Z^2/n^2$~for an arbitrary~$n)$.

    Adopting the `orthodox' point of view, Pauli now calculates the relativistic correction
    on the anomalous Zeeman effect, using his earlier results -- in particular his sum
    rule. I do not have to tell you this in detail, because it turns out that this
    influence on the $g$-factors is \textit{not} compatible with experience. The empirical
    factors $g$ are rational numbers depending only on the quantum numbers of the term. The
    result is summarized by Pauli as follows:
    \begin{quote}
    \textit{ ``In order to explain the observed factors $g$ by means of an angular momentum of
    closed shells, such as the K-shell of the alkali atoms, one would have to assume a
    doubling of the ratio of magnetic to mechanical momentum for electrons in the shell,
    and also a compensation of the classically computed relativistic effect of velocity,''}
    \end{quote}

    Pauli rejects this logical possibility. Instead he assumes that closed shells have no
    angular momentum and no magnetic moment. This implies that in the case of alkali atoms
    the angular momentum of the atom and its change of energy in a magnetic field are \textbf{due
    to the valence electron only}. In Pauli's words:
    \begin{quote}
    \textit{``Insbesondere werden bei den Alkalien die Impulswerte des Atoms und seine
    Energie\"anderungen in einem \"ausseren Feld im wesentlichen als alleinige Wirkung des
    Leuchtelektrons angesehen, das auch als Sitz der magneto-mechanischen Anomalie betrachtet
    wird.''}
    \end{quote}

    So far Pauli had only made a critical analysis of an existing hypothesis, but now comes
    a big jump when he writes:
    \begin{quote}
    \textit{``According to this point of view the doublet structure of alkali spectra as well as
    the deviation from Larmor's theorem is due to a particular two-valuedness of the
    quantum theoretical properties of the electron, which cannot be described from the
    classical point of view.''}
    \end{quote}

    Since Pauli does not explain these prophetic words any further in this second paper,
    it may be helpful if I add a few remarks. For \textit{strong} fields he had the
    formulae
    \[ M=M_L+M_S~,~~~ \Delta E/\mu_0B=M_L+2M_S~. \]

    Pauli follows Sommerfeld and interprets $M$ in his next paper as the total angular
    momentum in the direction of the field. (Sommerfeld also introduced the quantum number
    $J$.) For alkali atoms the closed shells do not contribute to $M$ nor to the magnetic
    moment. Hence,
    \[
    M=m_\ell+m_s~,~~~ \Delta E/\mu_0B = m_\ell+2m_s~, \]
    where $m_\ell, m_s$ are the values of $M_L$ and $M_S$ for the single valence electron.

    The integer number $m_\ell$ may be interpreted classically as the orbital angular
    momentum in the direction of the field. Therefore, $m_s$ is an \textit{intrinsic}
    contribution of the electron to the total angular momentum $M$ in the direction of the
    field which must be added to $m_\ell$. We have already seen that for the alkali
    doublets $m_s$ takes the values $\pm \frac 12$.

    Since $m_\ell$ is an integer it follows that $M$ is a half-integer, and since $J$ of a
    multiplet is defined to be the maximal value of $M$, we have for the \textit{two terms of an
    alkali doublet}
    \[
    J\;=\;L\;\pm\;1/2~. \]
    Thus, the two-valuedness of $J$, which is responsible for the doublet splitting, is a
    direct consequence of the two-valuedness of $m_s$. This explains the first part of
    Pauli's key sentence:
    \begin{quote}
    \textit{``...the doublet structure of alkali spectra (...) is due to a particular
    two-valuedness (...) of the electron.''}
    \end{quote}

    What exactly did Pauli mean concerning ``the deviation of Larmor's theorem''?

    In the paper of Pauli I have just discussed he uses the well-known formula for the
    energy of an atom in a magnetic field
    \[\triangle E = - \vec{M} \cdot \vec{B}~,~~~ \vec{M}~:~{\rm magn.~moment~.} \]
    If we compare this with his expression for strong fields, we see that an atom behaves
    in a strong field like a magnet having a magnetic moment $\mu_0 (M_L+2M_S)$ in the
    direction of the field. For a single valence electron this is equal to $\mu_0 (m_\ell + 2m_s)$.

    So far strong fields have been assumed. But if we consider an $S$ state of an alkali
    atom, we have $M_L=0$, $M=M_S=m_s$, and now the formula $\triangle E/\mu_0B = 2m_s$
    holds -- by Pauli's sum rule -- for \textit{weak} fields too. This means: For
    $S$ states the magnetic moment of alkali atoms is equal to $\underline{2m_s\mu_0}$. According
    to Pauli, this magnetic moment is \textit{entirely due to the valence electron}.

    Pauli did not attempt to give a meaning to the fourth degree of freedom in terms of a
    model. In his Nobel Prize lecture \cite{P6} he said about this:
    \begin{quote}
    \textit{``The gap was filled by Uhlenbeck and Goudsmit's idea of electron spin, which made it
    possible to understand the anomalous Zeeman effect. (...) Although at first I strongly
    doubted the correctness of this idea because of its classical mechanical character, I
    was finally converted to it by Thomas' calculations on the magnitude of doublet
    splitting.''}
    \end{quote}

    \vspace{.5cm}

    \noindent {\bf Step 3: The exclusion principle}

    \medskip

    \noindent

    In his decisive third paper \cite{P7} Pauli first summarizes his previous results for alkali
    metals. For these the quantum numbers $L,J,M$ of the atom coincide with those of the
    valence electron for which we use the modern notation $\ell, j, m_j$. (Pauli's notation
    is: ~$k_1=\ell+1,~ k_2=j+\,\frac 12 ~, m_1=m_j$.) Beside these there is, of course,
    also the principle quantum number $n$. As already explained, the number $j$ is equal to
    $j=\ell \,\pm\,\frac 1 2~.$ ~Pauli emphasizes:
    \begin{quote}
    \textit{``The number of states in a magnetic field for given $\ell$ and $j$ is $2j+1$,
    the number of these states for both doublets with a given $\ell$ taken together
    is ~$\underline{2(2\ell+1)}$.''}\end{quote}

    For the case of \textit{strong} fields, Pauli adds, one can use instead of $j$ the quantum number
    $m_2:=m_j\pm1/2~(=m_j+m_s)$, which directly gives the component of the magnetic moment parallel to
    the field. (We would use for the Paschen-Back region the four quantum numbers $n,~\ell,~m_{\ell},~ m_s$.)

    Next, Pauli extends the ``formal classification of the valence electron by the four
    quantum numbers $n, \ell, j, m_j$ to \textit{complicated atoms}''. This is performed
    with the help of Bohr's \textit{principle of permanence (Aufbauprinzip)}, which says:
    If, to a partially ionized atom, one (or more) electron is added, the quantum numbers of
    the electrons already bound remain the same as in the ionized atom. Pauli shows that in
    simple as well as in more complicated cases the application of this principle gives
    just the right variety of terms for the atom.

    For Pauli's further line of thought the formulae for the Zeeman effect in
    \textit{strong} fields are again essential. First, the principle of permanence implies
    that one can associate quantum numbers $m_j$ for the individual electrons, the sum of which
    is the total angular momentum of the atom in the direction of the field:
    \[
    M\;=\;\sum m_j~. \]
    By the same rule, the magnetic moment $(M+M_S)\mu_0$ is also equal to the sum
    of the moments $m_2\mu_0$ of all the electrons, i.e.,
    \[
    M_2 := M_L + 2M_S = M+M_S = \sum m_2~. \]
    In the sums $m_j$ and $m_2$ have to assume \textit{independently} all values which belong to the
    quantum numbers $j,\ell$. Pauli checked (for instance for neon) that this gives the
    correct results for the Zeeman terms.

    This result, he says, suggests the following hypothesis: \textit{``Every
    electron in the atom can be characterized by its principle quantum number $n$ and \textbf{three
    additional quantum numbers} $\ell, j, m_j $.''}. As for the alkali spectra, $j$ is always
    equal to $\ell\pm 1/2$. For strong fields the quantum number $m_2=m_j\pm 1/2$ is used instead of $j$.

    It must be emphasized that Pauli had to assume a magnetic field so strong that every
    electron has, \textit{independently} of the others, a definite mechanical angular momentum $m_j$ and a
    magnetic moment $m_2$ (in units of $\mu_0$), but he notes that for thermodynamic
    reasons (invariance of the statistical weights under adiabatic transformations of the
    system) the number of states in weak fields must be the same as in strong fields. In an
    article of van der Waerden \cite{W}, I have made heavy use of in this section, this is
    commented as follows:
    \begin{quote}
    \textit{``It is clear that the definition of these quantum numbers presented great difficulties
    at a time when quantum mechanics did not exist and the types of motion of the electron
    had to be described by inadequate classical models. (...) We have to admire Pauli's
    courage and persistence in developing the logical consequences of his hypothesis. The
    subsequent development of quantum mechanics led to a complete justification of every
    one of his assumptions.''}
    \end{quote}

    Next, Pauli considers the case of \textit{equivalent electrons}. First of all he notes
    that in this case some combinations of quantum numbers do \textit{not occur in nature}.
    For instance, if two valence electrons are in $s$ states belonging to different values
    of $n$, we observe a singlet $S$ term and a triplet $S$ term. If, however, both
    electrons have the same $n$, \textit{only the singlet term occurs}. For Pauli the
    question arises, which quantum theoretical rules govern this behavior of the terms.

    This reduction of terms, Pauli says, is closely connected with the phenomenon of closed
    shells. About this E.C. Stoner \cite{S} had recently made a new proposal which deviated from
    Bohr's theory of the periodic system. For example, Bohr had divided the 8 electrons of
    the $L$-shell into two subgroups of 4 electrons. Stoner, on the other hand, proposed to
    divide the electrons into a subgroup of 2 electrons having $\ell=0$, and a subgroup of 6
    electrons with $\ell=1$. Generally, for any closed shell and every value of $\ell < n$,
    Stoner associated a \textit{subgroup of $2(2\ell+1)$ electrons}.

    Even more important was Stoner's remark that the same number $2(2\ell+1)$ is also equal
    to the number of states of an \textit{alkali atom in a magnetic field} belonging to the same
    value of $\ell$ and to a given principle quantum number of the valence electron. This
    remark of Stoner gave Pauli the clue to his exclusion principle. He explains the fact
    that there are exactly $2(2\ell+1)$ electrons in every subgroup of a closed shell by
    assuming that every state, characterized by the quantum numbers $(n, \ell, j, m_j)$, is
    occupied by \textbf{just one electron}. Then we have for a given $n$ and $\ell >0$ just
    the two possibilities $j=\ell \pm \frac 1 2$ with $2j+1$ values for $m_j$, giving
    together $2(2\ell+1)$ electrons.

    In Pauli's words of his Nobel Prize lecture \cite{P6}:
    \begin{quote}
    \textit{``The complicated numbers of electrons in closed subgroups reduce to the simple
    number} \textbf{one} \textit{if the division of the groups by giving the values of the 4 quantum
    numbers of an electron is carried so far that every degeneracy is removed. A single
    electron already occupies an entirely non-degenerate energy level.''}
    \end{quote}

    In his \textit{original paper} Pauli enunciates his principle as follows:
    \begin{quote}
    \textit{``There can \textbf{never be two or more equivalent electrons} in an atom, for which in strong
    fields the values of all quantum numbers ~$n,\ell,j,\\ m_j$~ are the same. If an electron is
    present in the atom, for which these quantum numbers have definite values, this state
    is `occupied'.''}
    \end{quote}

    From this Pauli deduces the numbers \textbf{2, 8, 18, 32,...} of electrons in closed shells, and
    the \textbf{reduction of terms} for equivalent electrons. Several further applications are
    always in accordance with experience.

    At the end of his paper Pauli expresses the hope that a deeper understanding of quantum
    mechanics might lead to a derivation of the exclusion principle from more fundamental
    hypothesis. To some extent this hope was fulfilled in the framework of relativistic
    quantum field theory. Pauli's key role in establishing the \textit{spin-statistic theorem} is
    well-known (see, e.g., \cite{N1}).

    Initially Pauli was not sure to what extent his exclusion principle would hold good. In a
    letter to Bohr of 12 December 1924 Pauli writes \textit{`The conception, from which I start,
    is certainly nonsense. (...) However, I believe that what I am doing here is no greater
    nonsense than the hitherto existing interpretation of the complex structure. My nonsense is
    conjugate to the hitherto customary one.'} The exclusion principle was not immediately accepted,
    although it explained many facts of atomic physics. A few days after the letter to Bohr,
    Heisenberg wrote to Pauli on a postcard: \textit{`Today I have read your new work, and it is
    certain that I am the one who}  rejoices most \textit{about it, not only because you push
    the swindle to an unimagined, giddy height (by introducing}  individual \textit{electrons
    with 4 degrees of freedom) and thereby have broken all hitherto existing records of
    which you have insulted me. (...).'}.

    For the letters of Pauli on the exclusion principle, and the reactions of his influential
    colleagues, I refer to Vol.1 of the \textit{Pauli Correspondence}, edited by Karl von Meyenn
    \cite{M}. Some passages are translated into English in the scientific biography by
    Charles Enz \cite{Enz}.

    \subsection*{Exclusion principle and the new quantum mechanics}

    On August 26, 1926, Dirac's paper containing the Fermi-Dirac distribution was
    communicated by R. Fowler to the Royal Society. This work was the basis of Fowler's
    \textit{theory of white dwarfs}. I find it remarkable that the quantum statistics of identical
    spin-1/2 particles found its first application in astrophysics. Pauli's exclusion
    principle was independently applied to \textit{statistical thermodynamics} by Fermi.
    \footnote{According to Max Born, Pascual Jordan was actually the first who discovered
    what came to be known as the Fermi-Dirac statistics. Unfortunately, Born, who was editor
    of the \textit{Zeitschrift f\"ur Physik}, put Jordans paper into his suitcase when
    he went for half a year to America in December of 1925, and forgot about it. For further
    details on this, I refer to the interesting article \cite{Sch} by E.L. Schucking.} In the same
    year 1926, Pauli simplified Fermi's calculations, introducing the grand canonical
    ensemble into quantum statistics. As an application he studied the behavior of a gas in
    a magnetic field (paramagnetism).

    Heisenberg and Dirac were the first who interpreted the exclusion principle in the
    context of Schr\"odinger's wave mechanics for systems of more than one particle. In
    these papers it was not yet clear how the spin had to be described in wave mechanics.
    (Heisenberg speaks of spin coordinates, but he does not say clearly what he means by
    this.) The definite formulation was soon provided by Pauli in a beautiful paper \cite{P8},
    in which he introduced his famous \textit{spin matrices}.

    At this point the foundations of non-relativistic quantum mechanics had been completed
    in definite form. For a lively discussion of the role of the exclusion principle in physics
    and chemistry from this foundational period, I refer once more to the address \cite{E}
    of Ehrenfest.

    \section*{Part II. Stability of matter in bulk}

    One of the immediate great qualitative successes of quantum mechanics was that it
    implies the stability of atoms. A much less obvious consequence of the theory
    is that ordinary matter in bulk, held together by Coulomb forces, is also stable. The
    mystery of this fact before the dawn of quantum mechanics was described by Jeans in
    1915 with the following words \cite{J}:
    \begin{quote}
    \textit{``There would be a very real difficulty in supposing that the (force) law $1/r^2$ held
    down to zero values of $r$. For the force between two charges at zero distance would be
    infinite; we should have charges of opposite sign continually rushing together and,
    when once together, no force would be adequate to separate them (...).Thus matter in
    the universe would tend to shrink into nothing or to diminish indefinitely in size.''}
    \end{quote}
    In quantum mechanics the electrons cannot fall into the nuclei.

    \section{Stability of atoms and `ordinary' matter in bulk}

    Atoms and `ordinary' matter in bulk, consisting of a system of $N$ electrons and $k$
    nuclei with charges $Z_1e, ..., Z_ke$, can be well described by the mutual Coulomb
    interactions. For the discussion that follows we use the Hamiltonian
    \begin{equation}
     H = T_e + V_{eK} + V_{ee} +  V_{KK}.
     \end{equation}
    $T_e$ is the kinetic energy of the electrons, and the three potential energies
    $V_{eK},~V_{ee},~V_{KK}$ are the Coulomb energies between the electrons and nuclei,
    among the electrons and among the nuclei, respectively. We treat the nuclei as
    infinitely heavy in fixed positions $\mathbf{R}_1,...,\mathbf{R}_k$
    (Born-Oppenheimer approximation). Since we
    are mainly interested in lower bounds of the ground state energy of the system, this is
    not a serious simplification; if the nuclei are treated dynamically,
    the nuclear kinetic energy adds  a positive contribution.

    Two different notions of stability are useful.

    (i) \textit{Stability of the first kind}:
    \begin{equation}
     E(N,k,\underline{R}): = \inf_{\psi}~(\psi,H\psi)
    \end{equation}
    is finite for every $N,k$, and positions $\underline{R} = (\mathbf{R}_1,...,\mathbf{R}_k)$ of the
    nuclei.

    (ii) \textit{Stability of the second kind}: Assuming that $Z_j\leq Z$ for all
    $j=1,...,k$, then
    \begin{equation}
     E(N,k): = \inf_{\underline{R}}~E(N,k,\underline{R}) \geq-A(Z)(N+k),
    \end{equation}
    where $A$ depends only on $Z$.

    Four decades after non-relativistic quantum mechanics was developed, Dyson and
    Lenard gave the first rigorous proof of
    the  stability of the second kind for matter in bulk \cite{D1}. For this the Pauli principle for
    the electrons is essential, while the statistics of the nuclei does not matter. How crucial the
    exclusion principle really is, was demonstrated shortly afterwards by Dyson \cite{D2}.
    With the help of the variational principle, using a strongly correlated trial wave
    function, Dyson established the following inequality for \textbf{bosons}:

    If, without loss of generality $N\leq k$, then the ground state energy is bounded as
    \begin{equation}
     E(N,k)\leq -const~N^{7/5} Ry
    \end{equation}
    for large $N$. For Dyson's trial wave function, that was suggested by the work
    of Bogolubov on superfluidity and is related to the
    wave function used by Bardeen, Cooper and Schrieffer in their work on
    superconductivity, the constant in (4) is small ($\sim10^{-6}$). A satisfactory upper
    bound was very recently achieved in \cite{Sol}.

    Two decades later, it was shown by Conlon, Lieb and H-T. Yau \cite{L2} that the ground state
    energy can also be bounded from below as
    \begin{equation}
     E(N,k)\geq -AN^{7/5} Ry ,~
    \end{equation}
    with $A\simeq 0.2$ for $N=k$. The exact value of $A$ has recently been derived in
    \cite{LS}, and comes out of a mean field equation predicted by Dyson in \cite{D2}.

    So the ground state energy is expected to be close to $-N^{7/5}~Ry$. For $N\sim
    10^{23}$ this corresponds to a binding energy of $\sim 10^{32}~Ry~\sim 1~ megaton$. Thus
    the energy that would be released if two pieces of such bosonic matter with $N\sim
    10^{23}$ would be put together would be that of a hydrogen bomb; very explosive stuff
    indeed.

    \subsection{Stability of atoms}

    The stability of the first kind for isolated atoms is obvious, even without the Pauli
    principle for electrons. The remarkable fact is that the exclusion principle guarantees
    stability of the second kind. This can easily be demonstrated.

    For a single atom the Hamiltonian is
    \begin{equation}
     H_N = \frac{1}{2m}\sum_{i=1}^N \mathbf{p}_i^2 - Ze^2\sum_{i=1}^N
    \frac{1}{\mid\mathbf{x}_i\mid} + e^2\sum_{i<j}\frac{1}{\mid\mathbf{x}_i - \mathbf{x}_j\mid}~.
    \end{equation}
    Since the last term gives a positive contribution to the ground state energy, a lower
    bound for the ``unperturbed'' Hamiltonian $H_0$ (without the mutual Coulomb repulsion)
    gives a rough lower bound for $H_N$. But the ground state energy for $H_0$ is obtained
    by filling up the Balmer levels. The last completely filled level has principal quantum
    number $n_0$ determined by
    \[ 2\sum_{n=1}^{n_0} n^2 \leq N \leq2\sum_{n=1}^{n_0+1}n^2~, \]
    i.e.,
    \[ \frac{2}{3}n_0(n_0+\frac 12)(n_0+1)\leq N \leq \frac{2}{3}(n_0+1)(n_0+\frac32)(n_0+2)~.  \]

    The ground state energy $E_0$ of the unperturbed Hamiltonian satisfies in units of
    $Z^2Ry$
    \[ -\sum_{n=1}^{n_0+1} \frac{2n^2}{n^2}\leq E_0 \leq -\sum_{n=1}^{n_0}
    \frac{2n^2}{n^2}~,\]
    i.e. $-2(n_0+1)\leq E_0\leq -2n_0$. For large $N,~n_0=(\frac{3N}{2})^{1/3 }+ O(1)$, thus
    the ground state energy $E_N$ of $H_N$ is bounded as
    \begin{equation}
     E_{N}  \geq -2\Bigl(\frac{3}{2}\Bigr)^{1/3}N^{1/3}(1+O(N^{-1/3}))Z^2~Ry~.
    \end{equation}
    This inequality implies stability of the second kind. For a neutral atom this lower
    bound is proportional to $Z^{7/3}$.

    It is not difficult to derive also a \textit{upper bound} proportional to $N^{1/3}Z^2$,
    using the variational principle with the Slater determinant belonging to the shell state
    considered above. Using also the fact that the exchange term is non-positive, as well
    as the virial theorem for the direct Coulomb term, one easily finds
    \begin{equation}
     E_N \leq -2\Bigl(\frac{3}{2}\Bigr)^{1/3}\Bigl(1-\frac{N}{2Z}\Bigr)N^{1/3}Z^2\Bigl
    (1+O(N^{-1/3})\Bigr)~Ry~.
    \end{equation}
    These bounds can be improved.

    We note that the \textit{Thomas-Fermi theory} gives for neutral atoms
    \begin{equation}
    E_N^{TF}=-1.5375~Z^{7/3}~Ry~.
    \end{equation}

    \subsection{Size of large atoms}

    We are interested in an inequality for the size of an atom in its ground state, defined as
    \[ r:=\Bigl\{~\frac{1}{N}\Bigl\langle\sum_{i=1}^N \mathbf{x}_i^2\Bigr\rangle~\Bigr\}^{1/2}~,  \]
    where the angular bracket denotes the ground state expectation value.
    One expects, for instance on the basis of the Thomas-Fermi theory, that $r>const~N^{-1/3}$.

    As a first ingredient we use the following operator inequality ($\hbar=1$):
    \[ \frac 12\sum_{i=1}^N(\mathbf{p}_i^2 +\omega^2\mathbf{x}_i^2) \geq \omega
    N^{4/3}\frac{3^{4/3}}{4}\Bigl(1+O(N^{-1/3})\Bigr)~. \]
    This is obtained as the previous inequalities for $E_0$ of $H_0$, using that the energy
    levels of an isotropic harmonic oscillator are = $\frac 3 2 +n\omega$, with degeneracies
    $g_n=2\cdot\frac 1 2 (n+1)(n+2)$.

    Taking now the expectation value of this inequality with the ground state of $H_N$, and
    setting
    \[ \omega = \frac{4}{3^{4/3}N^{4/3}}\Bigl\langle\sum_{i=1}^N \mathbf{p}_i^2\Bigr\rangle
    \]
    leads, up to $N^{-1/3}$ corrections, to
    \[ \Bigl\langle\sum_{i=1}^N \mathbf{x}_i^2\Bigr\rangle \geq\frac{(3N)^{8/3}}
    {16 \Bigl\langle\sum_{i=1}^N \mathbf{p}_i^2\Bigr\rangle}~. \]
    Finally, we use the virial theorem for $H_N$ and the previous lower bound for $E_N$ to
    conclude that in units with $\hbar^2/2m = 1, e=1$
    \[ \frac 1 2\Bigl\langle\sum_{i=1}^N \mathbf{p}_i^2\Bigr\rangle =
    \mid E_N\mid\leq\Bigl(\frac{3}{2}\Bigr)^{1/3}N^{7/3}\Bigl(1+O(N^{-1/3}\Bigr)~.\]
    Together this gives
    \[ \Bigl\langle\sum_{i=1}^N \mathbf{x}_i^2\Bigr\rangle \geq
    \frac{9\cdot6^{1/3}}{32}N^{1/3}~, \]
    i.e.
    \begin{equation}
     r\geq 0.71~ N^{-1/3}\Bigl(1+O(N^{-1/3}\Bigr) ~.
    \end{equation}
    \subsubsection*{Supplementary remarks}

    For matter in bulk it is not possible to arrive at energy estimates in such an explicit
    and elementary fashion as for (7) and (8), and one has to use more general methods. As
    a nice illustration of these, we show how one arrives at a quite accurate lower bound
    without solving a differential equation, but by making use of the following \textit{Sobolev
    inequality} \cite{SO} for any $\psi\in L^2(\textbf{R}^3)$:

    \begin{equation}
    (\psi,T_e\psi) = \parallel \nabla\psi \parallel_2^2~ \geq K_s \parallel \psi
    \parallel_6^2 = K_s\parallel \rho_\psi\parallel_3~,
    \end{equation}
    where $\rho_\psi := \mid\psi\mid^2$ and $K_s=3(\pi/2)^{4/3}\simeq 5.5$ (this numerical
    value is known to be optimal). This inequality (which is a special case of a whole
    class), allows us to bound the ground state energy of hydrogen like atoms as
    \begin{equation}
    E\geq \inf_{\rho}~\Bigl\{h(\rho)~:~\rho(\mathbf{x})\geq 0,~~ \int_{\textbf{R}^3}\rho~
    d^3x=1\Bigr\}~,
    \end{equation}
    with
    \begin{equation}
    h(\rho)=K_s \parallel\rho\parallel_3 -Z\int~\frac{\rho(\textbf{x})}{\mid
    \textbf{x}\mid}~d^3x~.
    \end{equation}
    It is straightforward (a nice exercise for students) to find the minimizing $\rho$, and
    to show that it gives the lower bound
    \[ E\geq -\frac{4}{3}Z^2~Ry~. \]
    This instructive calculation is from Lieb's review paper \cite{RMP}.

    One does not loose much by using an even weaker inequality, which has the advantage to
    be generalizable to many electron systems. This is obtained from (11) with the help of
    H\"older's inequality:
    \[\parallel fg\parallel_1 \leq\parallel f\parallel_p \parallel g\parallel_{p'}~~~~~~
    (\frac{1}{p}+\frac{1}{p'}=1,~~ p\geq1.)\]
    For $p=3,~p'=\frac 3 2$ this implies for a normalized $\rho$
    \[ \int\rho^{5/3}\leq \parallel\rho\parallel_3 \parallel\rho^{2/3}\parallel_{3/2} =
    \Bigl (\int\rho^3
    \Bigr)^{1/3}\Bigl(\int\rho\Bigr)^{2/3}=\Bigl(\int\rho^3\Bigr)^{1/3}~.\]
    Hence,
    \begin{equation}
    (\psi,T_e\psi) \geq K_1 \int_{\textbf{R}^3} \rho_\psi(\textbf{x})^{5/3}~d^3x
    \end{equation}
    for $K_1=K_s$, but $K_1$ can be improved to $K_1=9.57$. Instead of (13) we now have
    the simpler functional
    \begin{equation}
    h(\rho)=K_1\int\rho^{5/3}~d^3x - Z\int\frac{\rho}{\mid\textbf{x}\mid}~d^3x~,
    \end{equation}
    but the lower bound comes out only slightly worse.

    For antisymmetric $N$-electron wave functions, Lieb and Thirring \cite{L3} were able to
    generalize (14), where now $\rho_\psi$ is the one-particle density, normalized as
    $\int\rho_\psi = N$. With the help of this generalized inequality they were able to
    bound $(\psi,H_N \psi)$ in terms of the Thomas-Fermi energy functional\footnote{Lieb and Simon
    \cite{L4} showed much earlier that the Thomas-Fermi theory becomes exact in the limit
    $Z\rightarrow \infty$, with the number of nuclei fixed.}.

    \subsection{The Dyson-Lenard-Lieb-Thirring Theorem}

    I have already mentioned that Dyson and Lenard gave in 1967 a proof of the stability of
    matter in the sense of (3). This proof was long and involved a large number of
    estimates. Even in sharp estimations it is unavoidable that about a factor two is
    lost per page. For a total of 40 pages of the paper one would thus expect a loss of about
    $2^{40}\sim 10^{14}$, and this is what actually happened. In his preface to Lieb's Selecta
    \cite{L1} Dyson writes:
    \begin{quote}
    \textit{``Our proof was so complicated and so unilluminating that it stimulated
    Lieb and Thirring to find the first decent proof. Why was our proof so bad and why was
    theirs so good? The reason is simple. Lenard and I began with mathematical tricks and
    hacked our way through a forest of inequalities without any physical understanding.
    Lieb and Thirring began with physical understanding and went on to find the appropriate
    mathematical language to make their understanding rigorous. Our proof was a dead end.
    Theirs was a gateway to the new world of ideas collected in this book.''}
    \end{quote}

    \subsubsection*{Heuristic considerations}

    On a heuristic level it is easy to understand the stability of the second kind of
    Coulomb dominated matter. Consider a neutral system of $N$ electrons and $N_Z$
    nuclei with charge $Ze$ and mass $m_Z~ (m_Z\simeq Am_N,~m_N=~$nucleon mass). Screening
    effects reduce the effective interactions essentially to one between nearest neighbors.
    Thus the Coulomb potential energy is roughly (for bosons and fermions)
    \[ V_{Coul}\approx -N_Z\frac{(Ze)^2}{(R/N_Z^{1/3})}~,\]
    where $R$ is the dimension of the system. For the  kinetic energy we have $T\approx
    Np^2/2m$, where $p$ is the average momentum of the electrons. Roughly speaking, the
    Pauli principle allows at most one electron in a de Broglie cube $(\hbar/p)^3$, and
    thus $ p \geq N^{1/3}\hbar/R $. For the total energy of the system, we therefore obtain
    the approximate inequality -- including for later purposes also the Newtonian potential
    energy $-\frac 1 2 N_Z^2Gm_Z^2/R$ of the nuclei --,
    \begin{equation}
    E\geq N\frac{p^2}{2m}-\frac{1}{2}\Bigl(\frac{N}{Z}\Bigr)^2\frac{Gm_Z^2}{\hbar
    N^{1/3}}p- \frac{Ne^2Z^{2/3}}{\hbar}p~.
    \end{equation}
    The minimum of the right hand side is attained for the average electron momentum $p_0$,
    given by
    \begin{equation}
     Np_0/m =\frac{1}{2}\Bigl(\frac{N}{Z}\Bigr)^2\frac{Gm_Z^2}{\hbar
    N^{1/3}}+ \frac{Ne^2Z^{2/3}}{\hbar}~,
    \end{equation}
    in terms of which the ground state energy is $E_0\approx -Np_0^2/2m$. Ignoring
    the gravitational interaction, this is \textit{linear in $N$}:
    \begin{equation}
    E_0\approx -N\cdot Ry~.
    \end{equation}
    For the electron density $n_0\approx(p_0/\hbar)^3$ and the matter density $\rho_0$ we
    obtain, if $a_0$ denotes the Bohr radius,
    \begin{equation}
    n_0\approx Z^2/a_0^3,~~~~\rho_0\approx AZm_N/a_0^3\approx 10~g/cm^3.
    \end{equation}

    If we would treat the electrons as bosons, we would only have the restriction imposed by
    the uncertainty relation, $p\geq \hbar/R$, and instead of (18) we would obtain
    \begin{equation}
    E_0\approx -N^{5/3}\cdot Ry~~~~~~(bosons).
    \end{equation}
    This $N^{5/3}$ law was established rigorously by Lieb for \textit{fixed positions of the
    nuclei}. However, when the nuclei are also treated dynamically the $N^{7/5}$ law, discussed earlier,
    holds.

    \subsubsection*{Rigorous bound}

    Lieb and Thirring \cite{L3} have established the rigorous bound

    \begin{equation}
    E(N,k)\geq -const\cdot\Bigl\{N+\sum_{j}^k Z_j^{7/3}\Bigr\}~Ry ~,
    \end{equation}
    with a constant of about 20 instead of $10^{14}$ in the work of Dyson and Lenard. The main step
    of the proof consists in bounding the ground state energy in terms of the Thomas-Fermi
    functional. Instead of minimizing this functional, Lieb and Thirring used a theorem of Teller
    stating that \textit{atoms do not bind in Thomas-Fermi theory} (see \cite{L4}). In this way a
    lower bound in terms of a lower bound of the Thomas-Fermi functional \textit{for atoms} was
    obtained, for which a previous result of Lieb and Simon could be used.

    \section{Stability and instability of cold stars}

    Once gravity becomes important we can no more expect stability of the second kind,
    because of the purely attractive and long range character of the gravitational
    interaction. Let us begin with some heuristic considerations.

    `Newton begins to dominate Coulomb' when the last two terms in (16) become
    comparable, i.e., for the `critical' electron number
    \[ N_c\approx
    Z\Bigl(\frac{Z}{A}\Bigr)^3\alpha^{3/2}\Bigl(\frac{M_{Pl}}{m_N}\Bigr)^3~.\]
    Here $\alpha$ is the fine structure constant and $M_{Pl}$ the Planck mass. Numerically
    this is about the number of electrons in Jupiter.

    For $N\gg N_c$ we can neglect the Coulomb contribution in (16) and then obtain from (17)
    \[ p_0/mc \approx\frac 1 2 \Bigl(\frac{A}{Z}\Bigr)^2\frac{m_N^2}{M_{Pl}^2}N^{2/3}~.\]
    This shows that the electrons become relativistic for
    \begin{equation}
    N>N_r:=\Bigl(\frac{Z}{A}\Bigr)^3\Bigl(\frac{2M_{pl}}{m_N}\Bigr)^{3/2}~.
    \end{equation}
    Therefore we treat the electrons in (16) relativistically. In units with $c=1$ we then
    have

    \begin{equation}
    E_0(N) \approx \inf_{p}~\Bigl\{ N \sqrt{p^2 + m^2}- \frac{1}{2} \Bigl(\frac{N}{Z}\Bigr)^2
    \frac{G m_Z^2}{\hbar N^{1/3}}~p\Bigr\} .
    \end{equation}
    One readily sees that \textit{the minimum only exists} for $N<N_r$. The corresponding
    limiting mass
    \begin{equation}
    M_r = (N_r/Z)m_Z\approx 2.8 \Bigl(\frac{Z}{A}\Bigr)^2\frac{M_{Pl}^3}{m_N^2}
    \end{equation}
    is close to the \textit{Chandrasekhar mass}.

    The delayed acceptance of the discovery by the 19 year old Chandrasekhar that quantum
    theory plus special relativity imply the existence of a limiting mass for white dwarfs
    belongs to the bizarre stories of astrophysics.

   The Fowler theory of white dwarfs is just the Thomas-Fermi theory, whereby the white
   dwarf is considered as a big ``atom'' with about $10^{57}$ electrons, and the
   Chandrasekhar theory is its relativistic version. In other words, the basic
   Chandrasekhar equation is the same as the relativistic Thomas-Fermi equation (for
   details see \cite{N3}). For white dwarfs the (semi-classical) Thomas-Fermi approximation
   is ideally justified (a rigorous result will be mentioned below).

   In this context the following close historical coincidence is interesting. Thomas' paper \cite{T}
   was presented at the Cambridge Philosophical Society on November 6, 1926. (Fermi's work was
   independent, but about one year later.) On the other hand, Fowler communicated his
   important paper \cite{F} on the non-relativistic theory of white dwarfs about one month later,
   on December 10, to the Royal Astronomical Society. I wonder who first noticed the close
   connection of the two approaches.

   It is worthwhile mentioning that Lieb and H-T. Yau have shown \cite{L5} that Chandrasekhar's
   theory can be obtained as a limit of a quantum mechanical description in terms of a
   semi-relativistic Hamiltonian.

   In a quantum mechanical description of white dwarfs the proper model to analyze  would be
   a Hamiltonian for electrons and ions with Coulomb and gravitational interactions, for
   which the kinetic energy of the electrons is the relativistic one:
   \begin{equation}
   T_e=\sum_{i=1}^N\Bigl[\sqrt{\mathbf{p}_i^2 + m^2}-m\Bigr]~.
   \end{equation}
   Unfortunately, a rigorous analysis of this model has (to my knowledge) not yet reached a
   satisfactory stage. A somewhat more modest goal has, however, been reached.

   The Coulomb forces in white dwarf material establish \textit{local} neutrality to a very
   high degree. For this reason the Coulomb interactions play energetically almost no role.
   (The corrections can be estimated and are on the few percent level.) The spatial
   distribution of the nuclei and hence their momentum distribution is much the same as
   those of the electrons. Based on these considerations it is reasonable to expect that
   the relevant Hamiltonian is \begin{equation}
   H_N=T_e - \sum_{1\leq i<j\leq N}\frac{G(m_Z/Z)^2}{\left|\mathbf{x}_i - \mathbf{x}_j\right|}
   \end{equation}
   ($m_Z/Z\simeq(A/Z)m_N$ is the mass in the star per electron).

   It is now natural to compare the ground state energy
   \[ E(N)= \inf_{\psi}~(\psi,H_N \psi) \]
   with the semi-classical energy of the Thomas-Fermi theory:
   \begin{equation}
   E^{TF}(N)= inf~\Bigl\{\mathcal{E}^{TF}(n)~:~\int n~=N\Bigr\}~,
   \end{equation}
   where
   \begin{equation}
    \mathcal{E}^{TF}(n)=\int_{\textbf{R}^3}\varepsilon(n(\mathbf{x}))~
    d^3x - \frac{\kappa}{2}\int~\frac{n(\textbf{x})n(\textbf{x'})}{\mid
    \textbf{x}-\textbf{x'}\mid}~d^3xd^3x'~.
    \end{equation}
    Here $\kappa=G(m_Z/Z)^2$ and
    \begin{equation}
    \varepsilon(n)=\frac{1}{\pi^2}\int_{0}^{p_F(n)}\Bigl[\sqrt{p^2+m^2}-m\Bigr]p^2dp~,
    \end{equation}
    where $p_F(n)$ is the Fermi momentum belonging to the number density $n$:
    \begin{equation}
    p_F(n)=(3\pi^2n)^{1/3}~.
    \end{equation}

    Lieb and H-T.Yau have proved the following\\
    \vspace{0.5cm}

    \textbf{Theorem}.~ \textit{Fix the quantity} $\tau=\kappa N^{2/3}$ \textit{at some value below the critical
    value} $\tau_c$ \textit{of the Chandrasekhar theory} ($\tau_c\simeq3.1$). \textit{Then}
    \begin{equation}
    \lim_{N\rightarrow \infty}E(N)/E^{TF}(N)=1.
    \end{equation}
    \textit{If} $\tau>\tau_c$, \textit{then}
    \begin{equation}
    \lim_{N\rightarrow \infty}E(N)=-\infty~.
    \end{equation}
    \vspace{0.5cm}

    As a corollary one can show that the ratio of the critical numbers of electrons for
    stability becomes 1 in the limit $G\rightarrow 0$.

    This demonstrates that we can study $H_N$ by means of the semi-classical approximation.
    This is, of course, not surprising. Indeed, corrections to the Thomas-Fermi
    approximations are of the order $N^{-1/3}$, i.e., of the order $10^{-19}$ for $N\sim
    10^{57}$. (In contrast to this tiny number for white dwarfs, corrections of the order
    $Z^{-1/3}$ for atoms are not negligible.)

    For an analogous discussion of \textit{boson stars}, I refer again to \cite{L5}; see
    also \cite{N2}.

   \section{Concluding remarks}

   For neutron stars such a quantum mechanical description is not possible, since general
   relativity has to be used. We are then bound to use a semi-classical description
   \`{a} la Thomas-Fermi, but from what has been said in the last section there can be
   no doubt that this is an excellent approximation.

   When GR is used as the correct theory of gravity, the exclusion principle still
   influences the magnitudes of limiting masses for stars. But
   while in Newtonian gravity theory the total energy of a system can be indefinitely
   negative, this is not the case in GR. The \textit{positive energy theorem} implies that it is
   impossible to construct an object out of ordinary matter, whose total energy is
   negative. (For a detailed proof and discussion, see, e.g., \cite{N3}). As a system is
   compressed to take advantage of the negative gravitational binding energy,
   a \textit{black hole} is eventually formed which has \textit{positive} total
   energy. This is, however, another story.

   \section*{Acknowledgements}

   I am very grateful to Elliott Lieb for his constructive criticism of the first version
   of this article. He also informed me on some recent work which I did not know.

    \end{document}